\documentclass[aps,prd,preprintnumbers,superscriptaddress,nofootinbib]{revtex4}%
\usepackage[dvipdfmx]{graphicx} 
\usepackage{feynmf}
\usepackage{color}
\usepackage{bm,latexsym,amsmath,amssymb,amsfonts,mathrsfs}
\usepackage[colorlinks=true,linkcolor=magenta,citecolor=magenta]{hyperref}

\usepackage{physics}	
\usepackage{xcolor}
\usepackage{tensor}
\usepackage{ulem}

\usepackage{physics}
\usepackage{aas_macros}

\newcommand*{\beqa}{\begin{eqnarray}}
\newcommand*{\eeqa}{\end{eqnarray}}

\newcommand*{\p}{\partial}
\newcommand{\realR}{\mathbb{R}}
\newcommand{\ra}{\rangle}
\newcommand{\la}{\langle}
\renewcommand{\H}{{\cal H}}

\newcommand{\A}{{\bf A}}
\newcommand{\B}{{\bf B}}
\renewcommand{\P}{{\bf P}}
\newcommand{\x}{{\bf x}}

\begin{document}

\title{Note on the Time Dilation of Charged Quantum Clocks}

\author{
Takeshi Chiba
}
\affiliation{Department of Physics, College of Humanities and Sciences, Nihon University, \\
                Tokyo 156-8550, Japan}
\author{
Shunichiro Kinoshita
}
\affiliation{Department of Physics, College of Humanities and Sciences, Nihon University, \\
                Tokyo 156-8550, Japan}
\affiliation{
Department of Physics, Chuo University, 
Kasuga, Bunkyo-ku, Tokyo 112-8551, Japan
}
\date{\today}

\begin{abstract}
We derive the time dilation formula for charged quantum clocks in electromagnetic fields.
As a concrete example of non-inertial motion, we consider a cyclotron motion in a uniform magnetic field.
Applying the time dilation formula to coherent state of the charged quantum clock, we evaluate the time dilation quantum-mechanically.
\end{abstract}


\maketitle

\section{Introduction}

Motivated by the tests of the weak equivalence principle in quantum regime, 
in our previous study we derived a formula of the averaged 
proper time read by one 
clock conditioned on another clock reading a different proper 
time  in a weak gravitational field \cite{Chiba:2022fzo} by 
extending the proper time observable proposed in \cite{Smith:2019imm}. 
The time dilation measured by these quantum clocks 
is found to have the same form as that in classical relativity. 
There, clocks are assumed to be in their inertial motion and 
their classical trajectories are the geodesics of 
the spacetime, that is, the clocks are always free-falling.  

Then, it would also be interesting to study what would happen for clocks 
in non-inertial motion. 
Any non-inertial motion should be caused by other external force than gravitational interaction, and it is not {\it a priori} clear whether 
the formalism in \cite{Smith:2019imm} can be extended to such situations. 
In classical relativity, we are allowed to assume a non-inertial trajectory without the equation of motion, so that we can evaluate its proper time kinematically.
On the other hand, in quantum theory we need to solve quantum dynamics to determine a trajectory.

In order to study the effect of non-inertial 
motion on the time dilation of quantum clocks, we consider charged 
quantum clocks interacting with the external electromagnetic fields 
as a concrete example. 
The study of a quantum charged particle  is also interesting 
in the light of quantum mechanics  in a rotating frame because 
there exists a close analogy between the motion in a rotating frame 
and the motion in a magnetic field  \cite{Sakurai:1980te}. 
Also,  a new class of optical clocks with highly charged ions 
has been received interest in recent years 
as references for highest-accuracy clocks and 
 precision tests of fundamental physics \cite{Kozlov:2018mbp,King:2022gtf}. 
 Such an optical  clock based on a highly charged ion 
 was recently realized \cite{King:2022gtf}.  Our study may be applicable to 
 such clocks.

 The paper is organized as follows.
 In Sec.~\ref{sec2}, we derive the time dilation formula for charged particles in electromagnetic fields and weak gravitational fields as the average of a proper time observable for a quantum clock. We extend the formalism 
 given in \cite{Smith:2019imm} to include the shift vector as well as the 
 electromagnetic field which is essential to treat a rotational motion and 
a rotating frame. 
 In Sec.~\ref{sec:time_dilation_cyclotron}, as a non-inertial motion, we consider the cyclotron motion in a uniform magnetic field.
 We evaluate the quantum time dilation by using the coherent state.
 In Appendix \ref{appendixA}, we summarize the several results of the coherent state for the cyclotron motion in quantum mechanics and the discussion of 
 the time dilation in a rotating frame.

\section{Charged Quantum Clock Particles in Spacetime}
\label{sec2}

\subsection{Classical Particles}

We consider a system of $N$ charged massive particles.  Each particle whose mass and charge are 
 $m_n$ and $q_n~(n=1,\dots,N)$ has a set of internal 
degrees of freedom, labeled by the configuration variables $\chi_n$ and 
their conjugate momenta $P_{\chi_n}$\cite{Smith:2019imm}. These internal 
degrees of freedom are supposed to represent the quantum clock.

The action of such a system in a curved spacetime with the metric 
$g_{\mu\nu}$ and an electromagnetic field $A_\mu$ is given by 
\beqa
S=\sum_n\int d\tau_n\left(-m_nc^2+q_nA_{\mu}\frac{dx_n^{\mu}}{d\tau_n}  +P_{\chi_n}\frac{d\chi_n}{d\tau_n}-
H_n^{\rm clock}\right) ,
\eeqa
 where $\tau_n$ is the proper time of the $n$th particle and 
 $H_n^{\rm clock}=H_n^{\rm clock}(\chi_n,P_{\chi_n})$ is a Hamiltonian 
 for its internal degrees of freedom.  
 
 Let  $x_n^{\mu}$ denote the spacetime position of the $n$th particle. 
 The trajectory of the $n$th particle $x_n^{\mu}(t)$  
 is parameterized by an arbitrary external time parameter $t$.  Noting that 
 $cd\tau_n=\sqrt{-g_{\mu\nu}\dot x_n^{\mu}\dot x_n^{\nu}}dt\equiv\sqrt{-\dot x^2_n}dt$, 
 where a dot denotes differentiation with respect to $t$,  
 the action is rewritten as 
 \beqa
 S=\int dt\sum_n\frac{1}{c}\sqrt{-\dot x_n^2}
 \left(-m_nc^2+q_nA_{\mu}\frac{\dot x_n^{\mu}c}{\sqrt{-\dot x_n^2}} +P_{\chi_n}\frac{\dot \chi_nc}{\sqrt{-\dot x_n^2}}-H_n^{\rm clock}\right) =: \int dt~ L .
 \label{action:L}
 \eeqa
 The momentum conjugate to $x_n^{\mu}$ is given by
 \beqa
 P_{n\mu}=\frac{\p L}{\p \dot x_n^{\mu}}=
 \frac{g_{\mu\nu}\dot x_n^{\nu}}{c\sqrt{-\dot x_n^2}}\left(m_n c^2+H_n^{\rm clock}\right) +q_nA_{\mu}.
 \eeqa
 Then the Hamiltonian associated with the Lagrangian $L$ is constrained to vanish:
\beqa
H=\sum_n\left(P_{n\mu}\dot x_n^{\mu}+P_{\chi_n}\dot \chi_n\right)-L\approx 0 .
\eeqa
In terms of the momentum, the constraints can be expressed in the form
\beqa
C_{H_n}:=g^{\mu\nu}(P_{n\mu}-q_nA_{\mu})(P_{n\nu}-q_nA_{\nu})c^2+\left(m_nc^2+H_n^{\rm clock}\right)^2\approx 0 .
\eeqa
Using the $(3+1)$ decomposition of the metric in terms of the lapse function $\alpha$, 
the shift vector $\beta^i$ and the three-metric $\gamma_{ij}$ such that \cite{mtw} 
\beqa
ds^2=-\alpha^2 c^2dt^2+\gamma_{ij}(dx^i+\beta^icdt)(dx^j+\beta^jcdt),
\label{3+1:metric}
\eeqa
the constraint is factorized in the form
\beqa
C_{H_n}&=&-\alpha^{-2}\left(P_{n0}-q_nA_0-\beta^i\left(P_{ni}-q_nA_i\right)\right)^2c^2+\gamma^{ij}(P_{ni}-q_nA_i)(P_{nj}-q_nA_j)c^2+\left(m_nc^2+H_n^{\rm clock}\right)^2\nonumber\\
&=& -\alpha^{-2}C_n^+C_n^-\approx 0 ,
\label{constraint-H}
\eeqa
where $C_n^{\pm}$ is defined by 
\beqa
C_n^{\pm}&:=&\left(P_{n0}-q_nA_0-\beta^i\left(P_{ni}-q_nA_i\right)\right)c\pm h_n ,
\label{constraint-pm}\\
h_n&:=&\alpha\sqrt{\gamma^{ij}(P_{ni}-q_nA_i)(P_{nj}-q_nA_j)c^2+(m_nc^2+H_n^{\rm clock})^2} .
\label{hamiltonian-n}
\eeqa
Note that we have set $x^0 = ct$.
Hereafter we assume that the spacetime is stationary.
The coordinates $x^{\mu}_n$ and their conjugate momenta $P_{n\mu}$ satisfy 
the fundamental Poisson brackets: $\{x_m^{\mu},P_{n\nu}\}=\delta_{mn}\delta^{\mu}_{\nu}$. 
The canonical momentum $P_{n\mu}$ generates translations in the spacetime coordinate $x_n^{\mu}$. 
Therefore, if $C_n^{\pm}\approx 0$, 
then $\pm h_n - q_n A_0 c-\beta^ic\left(P_{ni}-q_nA_i\right)$ is the generator of translation in the $n$th particle's time coordinate
and is the Hamiltonian for both 
the external and internal degrees of freedom of the $n$th particle.

\subsection{Quantization}

We canonically quantize the system of $N$ particles by promoting the phase space 
variables to operators acting on appropriate Hilbert spaces: 
$x^0_n$ and $P_{n0}$ become canonically conjugate self-adjoint operators acting on the Hilbert space ${\cal H}^0_n\simeq L^2(\mathbb{R})$ associated with 
the $n$th particle's temporal degree of freedom; operators $x_n^i$ and $P_{ni}$ 
acting on the Hilbert space 
${\cal H}_n^{\rm ext}\simeq L^2(\realR^3)$ associated with the particle's 
external degrees of freedom; operators $\chi_n$ and $P_{\chi_n}$ acting  
on the Hilbert space 
${\cal H}_n^{\rm clock}$ associated with the particle's 
internal degrees of freedom. Then the Hilbert space describing 
the $n$th particle is 
${\cal H}_n\simeq {\cal H}^0_n\otimes{\cal H}_n^{\rm ext}\otimes{\cal H}_n^{\rm clock}$.

The constraint equations (\ref{constraint-H}) now become operator equations restricting the physical state of the theory,  
\beqa
C_n^+C_n^-|\Psi\rangle\rangle=0,~~~~\forall n ,
\eeqa 
where $|\Psi\rangle\rangle\in {\cal H}_{\rm phys}$ is a physical state of a 
clock  $C$ and a system $S$ and lives in the 
physical Hilbert space ${\cal H}_{\rm phys}$. 

To specify ${\cal H}_{\rm phys}$, 
the normalization of the physical state in ${\cal H}_{\rm phys}$ is performed by 
projecting a physical state $|\Psi\rangle\rangle$ onto a subspace in which 
the temporal degree of freedom of each particle (clock $C$) 
is in an eigenstate $|t_n\rangle$ 
of the operator $x_n^0$ associated with the eigenvalue $t\in \realR$ in 
the spectrum of $x_n^0$: $x_n^0|t_n\rangle= c t|t_n\rangle$. 
The state of $S$ by conditioning $|\Psi\ra\ra$ on $C$ reading the time $t$ is then 
given by
\beqa
|\psi_S(t)\ra=\la t|\otimes I_S|\Psi\ra\ra ,
\eeqa
where $|t\ra=\otimes_n|t_n\ra$ and $I_S$ is the identity on 
${\cal H}\simeq \bigotimes_n\H_n^{\rm ext}\otimes\H_n^{\rm clock}$.  
We demand that the state  $|\psi_S(t)\ra$ is normalized as $\la\psi_S(t)|\psi_S(t)\ra=1$ for $\forall t\in \realR$  on 
a spacelike hypersurface defined by all $N$ particles' temporal degree 
of freedom being in the state $|t_n\ra$. The physical state $|\Psi\ra\ra$ 
is thus normalized with respect to the inner product~\cite{Smith:2019imm}:
\beqa
\la\la\Psi|\Psi\ra\ra_{PW}:=\la\la\Psi||t\ra\la t|\otimes I_S|\Psi\ra\ra
=\la\psi_S(t)|\psi_S(t)\ra=1, 
\eeqa
 and the physical state $|\Psi\ra\ra$ can be written as
\beqa
|\Psi\ra\ra=\int dt |t\ra\la t|\otimes I_S|\Psi\ra\ra=\int dt|t\ra|\psi_S(t)\ra .
\label{entangled}
\eeqa

Hereafter, we consider physical states that satisfy 
$C_n^+|\Psi\rangle\rangle=0$ for all $n \in \mathbb{N}$.
It can be shown  that the conditioned state $|\psi_S(t)\ra$ satisfies the 
Schr\"odinger equation with  $t$ as a time parameter \cite{Smith:2019imm}:
\beqa
i\hbar \frac{d}{dt}|\psi_S(t)\ra=H_S|\psi_S(t)\ra ,
\label{schrodinger}
\eeqa
where $H_S$ is given by
\beqa
H_S=\sum_n\left(h_n-q_nA_0c-\beta^ic\left(P_{ni}-q_nA_i\right)\right)\otimes I_{S-n}\equiv \sum_n\widetilde{h_n}\otimes I_{S-n}
\label{hamiltonian}
\eeqa
with $I_{S-n}$ being the identity 
on $\bigotimes_{m\neq n}\H^{\rm ext}_m\otimes\H^{\rm clock}_m$.  
Therefore,   $|\psi_S(t)\ra$ can be regarded as the 
time-dependent state of the $N$-particles with the Hamiltonian $H_S$ 
evolved with the external time $t$. 

\subsection{Probabilistic Time Dilation}
 
Consider two clock particles $\mathrm{A}$ and $\mathrm{B}$ with internal degrees of freedom.
Each clock is defined to be the quadrupole 
$\{\H^{\rm clock}_n,\rho_n,H^{\rm clock}_n, T_n\}$, 
where $\rho_n$ is a fiducial state and $T_n$ is proper time observable for 
$n\in\{\mathrm{A},\mathrm{B}\}$. 
The proper time observable is defined as a POVM (positive operator valued measure)
\beqa
T_n:=\left\{ E_n(\tau)~\forall \tau \in G ~{\rm s.t.}\int_G d\tau E_n(\tau)=I_n\right\} ,
\eeqa
where $E_n(\tau)=|\tau\ra\la\tau|$ is a positive operator on $\H^{\rm clock}_n$, $G$ is the group generated by $H^{\rm clock}_n$, and $|\tau\ra$ is a 
clock state associated with a measurement of the clock yielding the time $\tau$. 
 
To probe time dilation effects between two clocks, we consider 
the  probability that clock~$\mathrm{A}$ reads the proper time $\tau_\mathrm{A}$ 
conditioned on clock~$\mathrm{B}$ reading the proper time $\tau_\mathrm{B}$
\cite{Page:1983uc,1984IJTP...23..701W}.
This conditional probability is given in terms of the physical state as
\beqa
{\rm Prob}[T_\mathrm{A}=\tau_\mathrm{A}|T_\mathrm{B}=\tau_\mathrm{B}]=\frac{\la\la\Psi|E_\mathrm{A}(\tau_\mathrm{A})
E_\mathrm{B}(\tau_\mathrm{B})|\Psi\ra\ra}{\la\la\Psi|
E_\mathrm{B}(\tau_\mathrm{B})|\Psi\ra\ra} \,.
\label{conditional-prob}
\eeqa

Consider the case where two clock particles $\mathrm{A}$ and $\mathrm{B}$ are moving in a 
curved spacetime. 
Suppose that initial conditioned state is unentangled, 
$|\psi_S(0)\ra=|\psi_{S_\mathrm{A}}\ra |\psi_{S_\mathrm{B}}\ra$, and that the external and internal degrees of freedom of both particles are unentangled, 
$|\psi_{S_n}\ra=|\psi_n^{\rm ext}\ra |\psi_n^{\rm clock}\ra$. 
Then, from Eq. (\ref{entangled}), 
the physical state takes the form
\beqa
|\Psi\ra\ra=\int dt |t\ra |\psi_S(t)\ra
=\int dt \bigotimes_{n\in \{\mathrm{A},\mathrm{B}\}}e^{-i\widetilde{h}_nt/\hbar}|\psi_n^{\rm ext}\ra|\psi_n^{\rm clock}\ra\, ,
\eeqa 
where $\widetilde{h}_n$ is defined in Eq. (\ref{hamiltonian}).
Further suppose that $\H^{\rm clock}_n\simeq L^2(\realR)$ so that 
we may consider an ideal clock such that  
 $P_n = H^{\rm clock}_n/c$ and $c T_n$ are 
the momentum and position operators on  $\H^{\rm clock}_n$. 
The canonical commutation relation yields 
$[c T_n, P_n] = [T_n, H^{\rm clock}_n] = i\hbar$. 
Then, the clock states are orthogonal $\la \tau |\tau'\ra=\delta(\tau-\tau')$ and 
 satisfy the covariance relation 
$|\tau+\tau'\ra=e^{-iH^{\rm clock}_n\tau'/\hbar}|\tau\ra$.
The conditional probability  (\ref{conditional-prob}) becomes
\beqa
{\rm Prob}[T_\mathrm{A}=\tau_\mathrm{A}|T_\mathrm{B}=\tau_\mathrm{B}]=\frac{\int dt~{\rm tr}[E_\mathrm{A}(\tau_\mathrm{A})\rho_\mathrm{A}(t)]{\rm tr}[E_\mathrm{B}(\tau_\mathrm{B})\rho_\mathrm{B}(t)]}{\int dt~{\rm tr}[E_\mathrm{B}(\tau_\mathrm{B})\rho_\mathrm{B}(t)]} ,
\label{conditional-prob2}
\eeqa
where $\rho_n(t)$ is the reduced state of the internal clock degrees of freedom defined as 
\cite{Smith:2019imm}
\beqa
\rho_n(t)={\rm tr}_{\H_S\backslash\H_n^{\rm clock}}\left(e^{-iH_St/\hbar}|\psi_{S_n}\ra\la\psi_{S_n}|e^{iH_St/\hbar}\right)
\label{reduced state}
\eeqa
with the trace over the complement of the clock Hilbert space. 

We assume that the fiducial states of the internal clock degrees of freedom 
are  the Gaussian wave packets centered
 at $\tau=0$ with width $\sigma$:
\beqa
|\psi_n^{\rm clock}\ra=\frac{1}{\pi^{1/4}\sigma^{1/2}}\int d\tau~ e^{-\frac{\tau^2}{2\sigma^2}}|\tau\ra\,.
\label{clock state}
\eeqa

Note that in evaluating the conditional probability (\ref{conditional-prob2}) 
by using Eq. (\ref{reduced state}) and Eq. (\ref{clock state}),  the terms in the Hamiltonian $H_S$ (\ref{hamiltonian}) which involve both the clock Hamiltonian $H_n^{\rm clock}$ and the external degrees of freedom survive.  
Therefore,  as in our previous study \cite{Chiba:2022fzo}, the conditional probability depends only on $h_n$ defined in Eq. (\ref{hamiltonian-n}) and is independent of   
the terms  in Hamiltonian $H_S$ which depend only 
on the external degrees of freedom (such as $A_0$ and $\beta^i$).

\subsection{Time Dilation}

In order to find the coupling of the clock Hamiltonian $H_n^{\rm clock}$ and the external degrees of freedom, we expand  
$\widetilde{h_n}$ in the effective Hamiltonian (\ref{hamiltonian}) 
according to the power of 
$H_n^{\rm clock}$ assuming $ H_n^{\rm clock}\ll m_nc^2$
\beqa
\widetilde{h_n}&=&\alpha \sqrt{\gamma^{ij}(P_{ni}-q_nA_i)(P_{nj}-q_nA_j)c^2+
m_n^2c^4} -q_nA_0c-\beta^ic\left(P_{ni}-q_nA_i\right)\nonumber\\
&&+\frac{\alpha m_n^2c^4}{\sqrt{\gamma^{ij}(P_{ni}-q_nA_i)(P_{nj}-q_nA_j)c^2+
m_n^2c^4} }\frac{H_n^{\rm clock}}{m_nc^2}+O((H_n^{\rm clock}/m_nc^2)^2) .
\eeqa
The term in the second line which involves both the clock Hamiltonian $H_n^{\rm clock}$ and the external degrees of freedom is relevant 
in calculating the conditional probability. One may recognize that the 
coefficient of $H_n^{\rm clock}$ is (minus of) the kinetic term of the 
$n$-th particle in the Lagrangian (\ref{action:L}), 
that is, $m_nc\sqrt{-\dot x_n^2} = m_nc^2 d\tau_n /dt$. 
This implies that the average of the time dilation would be given by the same form as the classical time dilation formula in the leading order of the clock Hamiltonian.
In other words, regardless of inertial or non-inertial motions, the time dilation would be given by difference of the proper time and distance between trajectories of each particle.

As a concrete example, 
in the Newtonian approximation of spacetime, the metric is given by 
$g_{00}=-\alpha^2=-(1+2\Phi(\x)/c^2),\gamma_{ij}=\delta_{ij}$, and $\beta^i=0$, where $\Phi(\x)$ is the Newtonian gravitational potential. 
$\widetilde{h_n}$  is then further expanded according to 
the number of the inverse power of $c^2$ as
\beqa
\widetilde{h_n}
= m_nc^2+H_n^{\rm clock}+H_n^{\rm ext}+H_n^{\rm int}+O(c^{-4}),
\eeqa
where the rest-mass energy term $m_nc^2$ is a constant and can be disregarded in 
$h_n$. The external Hamiltonian $H_n^{\rm ext}$ and the interaction Hamiltonian 
$H_n^{\rm int}$ are given by
\beqa
H_n^{\rm ext}&:=&\frac{\delta^{ij}(P_{ni}-q_nA_{ni})(P_{nj}-q_nA_{nj})}{2m_n}+m_n\Phi_n-q_nA_{n0}c\equiv \frac{(\P_n-q_n\A_n)^2}{2m_n}+
m_n\Phi_n-q_nA_{n0}c,\label{hamiltonian:ext}\\
H_n^{\rm int}&:=&-\frac{(\P_n-q_n\A_n)^2H_n^{\rm clock}}{2m_n^2c^2}
+\frac{\Phi_nH_n^{\rm clock}}{c^2}
-\frac{1}{2m_nc^2}\left(\frac{(\P_n-q_n\A_n)^2}{2m_n}-m_n\Phi_n\right)^2+O(c^{-4})
\,,
\eeqa 
where $A_{n\mu}:=A_{\mu}(\x_n), \Phi_n:=\Phi(\x_n)$. 

The reduced state of the internal clock becomes 
\beqa
\rho_n(t)&=&{\rm tr}_{\H_S\backslash\H_n^{\rm clock}}\left[
e^{-iH_St/\hbar}|\psi_{S_n}\ra\la\psi_{S_n}|e^{iH_St/\hbar}\right]\nonumber\\
&=&\bar\rho_n(t)-it~ {\rm tr}_{\rm ext}\left(
[H_n^{\rm int}, \bar\rho_n^{\rm ext}(t)\otimes \bar\rho_n(t)]+O((H_n^{\rm int} t)^2)
\right)\nonumber\\
&=&\bar\rho_n(t)+it \left(\frac{\la (\P_n-q_n\A_n)^2\ra}{2m_n^2c^2} -\frac{\la \Phi_n\ra}{c^2}\right)
[H_n^{\rm clock},\bar\rho_n(t)]+O(c^{-4}),
\eeqa
where $\bar \rho_n(t)=e^{-iH_n^{\rm clock}t/\hbar}\rho_n e^{iH_n^{\rm clock}t/\hbar}$ and 
$\bar \rho_n^{\rm ext}(t)=e^{-iH_n^{\rm ext}t/\hbar}\rho_n^{\rm ext} e^{iH_n^{\rm ext}t/\hbar}$.
The conditional probability  (\ref{conditional-prob}) is evaluated to leading relativistic order as
 \beqa
{\rm Prob}[T_\mathrm{A}=\tau_\mathrm{A}|T_\mathrm{B}=\tau_\mathrm{B}]&=&
\frac{e^{-\frac{(\tau_\mathrm{A}-\tau_\mathrm{B})^2}{2\sigma^2}}}{\sqrt{2\pi}\sigma}\left[1+
\left(\frac{\la (\P_\mathrm{A}-q_\mathrm{A}\A_\mathrm{A})^2\ra}{4m_\mathrm{A}^2c^2}-\frac{\la (\P_\mathrm{B}-q_\mathrm{B}\A_\mathrm{B})^2\ra}{4m_\mathrm{B}^2c^2}-\frac{\la\Phi_\mathrm{A}\ra}{2c^2}+\frac{\la\Phi_\mathrm{B}\ra}{2c^2}
\right)\right.  \nonumber\\
&&~~~~~~~~~~~~~~~~~~~~\left.\times
\left(1-\frac{\tau_\mathrm{A}^2-\tau_\mathrm{B}^2}{\sigma^2}\right)
\right] ,
\eeqa
where $\la H_n^{\rm ext}\ra=\la\psi^{\rm ext}_n|H_n^{\rm ext}|\psi_n^{\rm ext}\ra$. 
Then the average proper time read by clock~$\mathrm{A}$ conditioned on clock~$\mathrm{B}$ indicating 
the time $\tau_\mathrm{B}$  is  
\beqa
\la T_\mathrm{A}\ra&=&\int d\tau~{\rm Prob}[T_\mathrm{A}=\tau|T_\mathrm{B}=\tau_\mathrm{B}]\tau\nonumber\\
&=&
\tau_\mathrm{B}\left[1-
\left(\frac{\la (\P_\mathrm{A}-q_\mathrm{A}\A_\mathrm{A})^{2}\ra}{2m_\mathrm{A}^2c^2}-\frac{\la\Phi_\mathrm{A}\ra}{c^2}\right)+\left(\frac{\la (\P_\mathrm{B}-q_\mathrm{B}\A_\mathrm{B})^{2}\ra}{2m_\mathrm{B}^2c^2}-\frac{\la\Phi_\mathrm{B}\ra}{c^2}\right)
\right]\,.
\label{timedelay:final}
\eeqa
This is the quantum analog of time dilation formula for the charged
 particles in the Newtonian gravity, extending the time dilation formula for neutral particles derived in \cite{Chiba:2022fzo}. 
 Noting that the time evolution of the position 
  $\dot \x_n=\frac{[\x_n,H_n^{\rm ext}]}{i\hbar}=\frac{\P_n-q_n\A_n}{m_n}$ from the Heisenberg equation,\footnote{Note that the equation becomes $\dot \x_n=\frac{\P_n-q_n\A_n}{m_n}-\boldsymbol{\beta}c$ in the presence of the shift 
  vector. } one may recognize that this 
  time dilation formula has the same form as the classical time 
  dilation in the Newtonian gravity. 
  The time dilation formula of Eq. (\ref{timedelay:final}) can also 
  be regarded as the extension of  the proper time observable proposed 
in \cite{Smith:2019imm} to non-inertial motion. 
  The time dilation of a clock, 
  regargless of  whether it is  in inertial motion or non-inertial motion,  
  is induced by its velocity and gravitational potential.

 \section{Time Dilation in a Uniform Magnetic Field}
 \label{sec:time_dilation_cyclotron}

As an application of the time dilation formula of Eq. (\ref{timedelay:final}),  
we consider the motion of a charged particle in a uniform magnetic field $B$ 
along the $z$ direction.  The quantum mechanics of the charged particle and the coherent state are discussed in detail in Appendix \ref{appendixA}. 
For the particle moving in the $xy$-plane in the flat spacetime, the Hamiltonian is given by
\beqa
H_n^{\rm ext}=\frac{({\bf P}_n-q_n\A_n)^2}{2m_n} .
\eeqa 
The time dilation formula (\ref{timedelay:final}) 
is reduced to the difference of the Hamiltonian 
\beqa
\la T_\mathrm{A}\ra=\tau_\mathrm{B}\left(1-\frac{\la H_\mathrm{A}^{\rm ext}\ra}{m_\mathrm{A}c^2}+\frac{\la H_\mathrm{B}^{\rm ext}\ra}{m_\mathrm{B}c^2}\right) .
\label{timedelay2}
\eeqa
Although  Eq. (\ref{timedelay2})  has the same form as the time dilation 
for neutral particles in inertial motion \cite{Smith:2019imm}, 
the interpretation is different: the former is the time dilation for particles in non-inertial motion while the latter is for particles in inertial motion. 
Since $H_n^{\rm ext}$ does not depend on the external time $t$ explicitly, 
the expectation value of $H_n^{\rm ext}$ is conserved. 
Therefore, the time dilation does not depend on $t$ 
in contrast to the gravitational time dilation \cite{Chiba:2022fzo}.
In the following, we calculate the time dilation between a charged 
quantum clock~$\mathrm{A}$ (with its charge $q_\mathrm{A}$) and 
an uncharged ($q_\mathrm{B}=0$)  quantum 
clock~$\mathrm{B}$  for coherent state. 
We also note that as explained in  Appendix \ref{appendixA}  
the time dilation formula (\ref{timedelay2}) 
does not change even if we move to a rotating frame. 

\subsection{Time Dilation in Coherent State}

It is known that the cyclotron motion of a
charged particle in a uniform magnetic field can be quantum mechanically well-described by the coherent state.
We consider the coherent state $|\alpha,\beta\ra$ defined by Eq. (\ref{coherentstate}) for the charged clock~$\mathrm{A}$.\footnote{$\alpha$ should not be confused with the lapse function in Eq. (\ref{3+1:metric}). }
Introducing the cyclotron frequency $\omega_\mathrm{A}=q_\mathrm{A}B/m_\mathrm{A}$ and  the radius of the cyclotron motion $r_0$, 
the center of the cyclotron motion $(X_0,Y_0)$ is related to $\beta$ as $X_0 - iY_0 = \sqrt{\frac{2\hbar}{m_\mathrm{A}\omega_\mathrm{A}}}\beta$ and the relative position $(r_0\cos\theta_0, r_0\sin\theta_0)$ is 
related to $\alpha$ as $r_0e^{i\theta_0}=\sqrt{\frac{2\hbar}{m_\mathrm{A}\omega_\mathrm{A}}}\alpha$. The expectation value of the position of 
the charged particle rotates clockwise with the angular velocity $\omega_\mathrm{A}$ 
about the center (see Eq. (\ref{mean:xi}) and Eq. (\ref{mean:eta})).  
Note that the uniform magnetic field $B$ is given by the vector potential 
$\A=\frac{B}{2}(-y,x,0)$ in the symmetric gauge.
{}From Eq. (\ref{hamiltonian:coherent}), 
the expectation value of the external Hamiltonian of the clock~$\mathrm{A}$ becomes 
\beqa
\la\alpha,\beta|H_\mathrm{A}^{\rm ext}|\alpha,\beta\ra=\hbar\omega_\mathrm{A} \left(|\alpha|^2+\frac12\right)=\frac12 m_\mathrm{A}\omega_\mathrm{A}^2r_0^2+\frac12\hbar\omega_\mathrm{A}\,.
\label{average:HA:coherent}
\eeqa

On the other hand, 
we assume that the state of the uncharged clock~$\mathrm{B}$ is a Gaussian state centered at  
$(x_\mathrm{B},y_\mathrm{B})=(x_{\mathrm{B}0},y_{\mathrm{B}0})$ with width $\sigma_\mathrm{B}$, whose wave function is 
\beqa
\la \x_\mathrm{B}|\psi_\mathrm{B}\ra=\left(\pi \sigma_\mathrm{B}^2\right)^{-1/2}\exp\left[-\frac{(x_\mathrm{B}-x_{\mathrm{B}0})^2+(y_\mathrm{B}-y_{\mathrm{B}0})^2}{2\sigma_\mathrm{B}^2}\right]\, .
\eeqa
Then, the expectation value of the external Hamiltonian of the clock~$\mathrm{B}$ becomes
\beqa
\la\psi_\mathrm{B}|H_\mathrm{B}^{\rm ext}|\psi_\mathrm{B}\ra=\frac{\hbar^2}{2m_\mathrm{B}\sigma_\mathrm{B}^2}\,.
\eeqa

Putting these together, the observed average time dilation between two clocks is given by
\beqa
\la T_\mathrm{A}\ra&=&\tau_\mathrm{B}\left(1-\frac{\la H_\mathrm{A}^{\rm ext}\ra}{m_\mathrm{A}c^2}+
\frac{\la H_\mathrm{B}^{\rm ext}\ra}{m_\mathrm{B}c^2}\right)\nonumber\\
&=&\tau_\mathrm{B}\left(1-\frac{\omega_\mathrm{A}^2r_0^2}{2c^2}-
\frac{\hbar\omega_\mathrm{A}}{2m_\mathrm{A}c^2}+\frac{\hbar^2}{2m_\mathrm{B}^2c^2\sigma_\mathrm{B}^2}\right)\,.
\eeqa

\subsection{Superposition}

\begin{figure}[htp]
	\centering
	\includegraphics[width=0.55\textwidth]{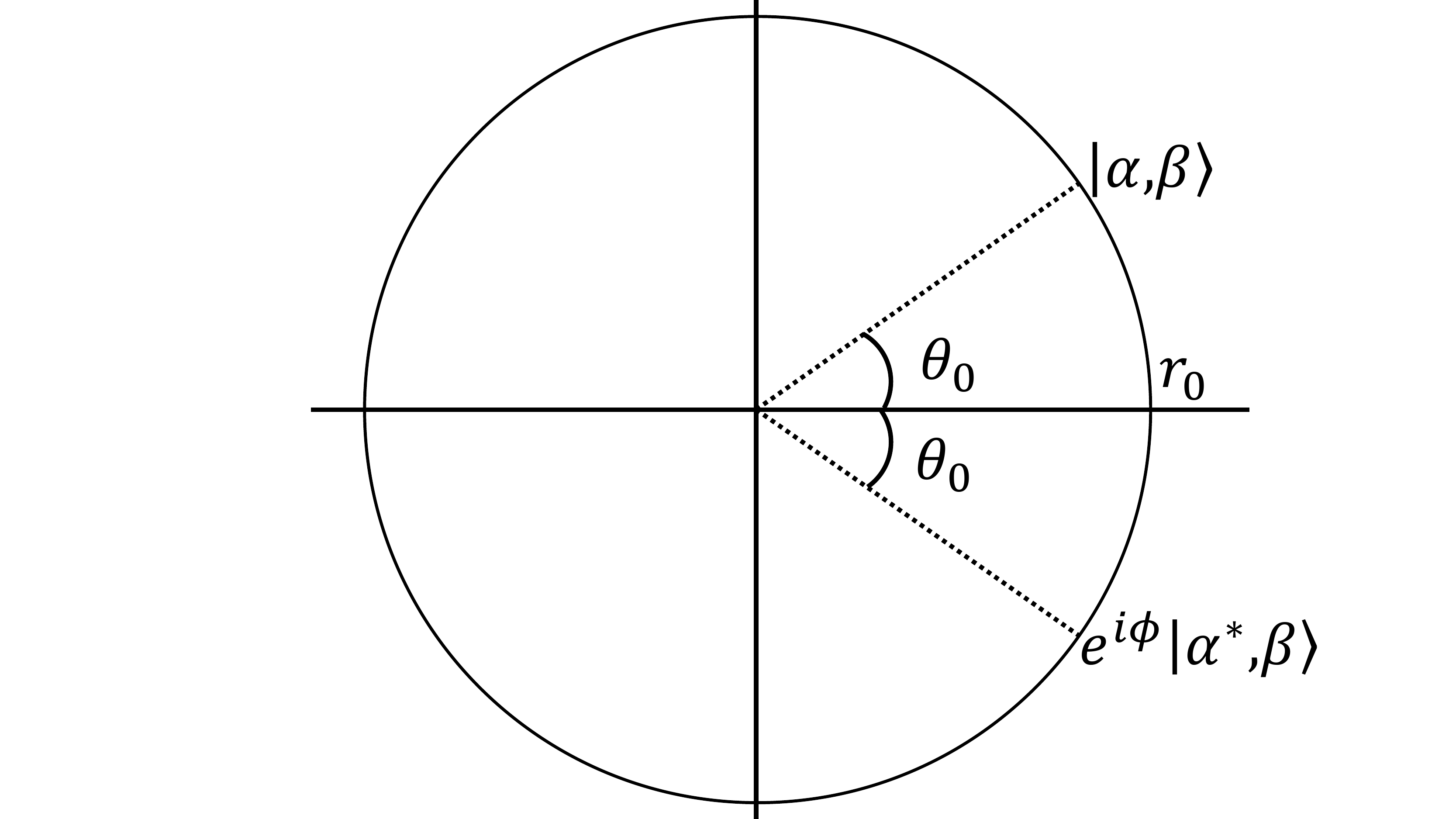}
	\caption{ \label{fig1}
	The superposition of two coherent states $|\alpha,\beta\rangle$ and $|\alpha^*,\beta\rangle$.  The radius of the circle is 
	$r_0=\sqrt{2\hbar/m_\mathrm{A}\omega_\mathrm{A}}|\alpha|$ and the angular separation is 
	$2\theta_0=2\tan^{-1}({\rm Im}~ \alpha/{\rm Re}~\alpha)$.
	 } 
\end{figure}

Next, we consider two clocks~$\mathrm{A}$ and $\mathrm{B}$ and suppose that initially clock~$\mathrm{A}$ is in a superposition of two coherent state \cite{1991PhRvA..44.2172S}:
\beqa
|\psi_\mathrm{A}\ra=\frac{1}{\sqrt{N}}\left(|\alpha,\beta\ra+e^{i\phi}|\alpha',\beta\ra\right)\,.
\eeqa
Two coherent states are assumed to have the same center 
of circle, namely the same $\beta$, but have different  positions on the circle as shown in Fig. \ref{fig1} :
\beqa
\alpha&=&\sqrt{\frac{m_\mathrm{A}\omega_\mathrm{A}}{2\hbar }}r_0e^{i\theta_0} ,\\
\alpha'&=&\sqrt{\frac{m_\mathrm{A}\omega_\mathrm{A}}{2\hbar }}r_0e^{-i\theta_0}=\alpha^*\, ,
\eeqa
which means the angular separation is $2\theta_0$ for $0 \le \theta_0 <\pi$. 
Two clocks rotate about the center clockwise with the angular velocity $\omega_\mathrm{A}$. 
The normalization factor $N$ is given by
\beqa
N=2+2{\rm Re}\left(e^{-i\phi}\la\alpha',\beta|\alpha,\beta\ra\right)=2+2{\rm Re}\left(e^{-i\phi}e^{\alpha^2-|\alpha|^2}\right) .
\eeqa
Then, the average of $H_\mathrm{A}^{\rm ext}$ is
\beqa
\la\psi_\mathrm{A}|H_\mathrm{A}^{\rm ext}|\psi_\mathrm{A}\ra&=&\hbar \omega_\mathrm{A}\left(|\alpha|^2+\frac12\right)
+\frac{2\hbar\omega_\mathrm{A}}{N}{\rm Re}\left((\alpha^2-|\alpha|^2)e^{-i\phi}e^{\alpha^2-|\alpha|^2}\right)\nonumber\\
&=&\frac12 m_\mathrm{A}\omega_\mathrm{A}r_0^2+\frac12\hbar\omega_\mathrm{A}+2\sin\theta_0
\frac{m_\mathrm{A}\omega_\mathrm{A}^2r_0^2}{N}{\rm Re}\left(ie^{i(\theta_0-\phi)}e^{\alpha^2-|\alpha|^2}\right) .
\eeqa
Hence the time dilation between two clocks becomes
\beqa
\la T_\mathrm{A}\ra
=\tau_\mathrm{B}\left(1-\frac{\omega_\mathrm{A}^2r_0^2}{2c^2}-
\frac{\hbar\omega_\mathrm{A}}{2m_\mathrm{A}c^2}-2\sin\theta_0
\frac{\omega_\mathrm{A}^2r_0^2}{Nc^2}{\rm Re}\left(ie^{i(\theta_0-\phi)}e^{\alpha^2-|\alpha|^2}\right)
+\frac{\hbar^2}{2m_\mathrm{B}^2c^2\sigma_\mathrm{B}^2}\right)\,.
\label{quantum:dilation}
\eeqa
The  term proportional to $\sin\theta_0$ arises from quantum interference  
due to the superposition and may be regarded as the quantum time dilation.

\begin{figure}[htp]
	\centering
	\includegraphics[width=0.80\textwidth]{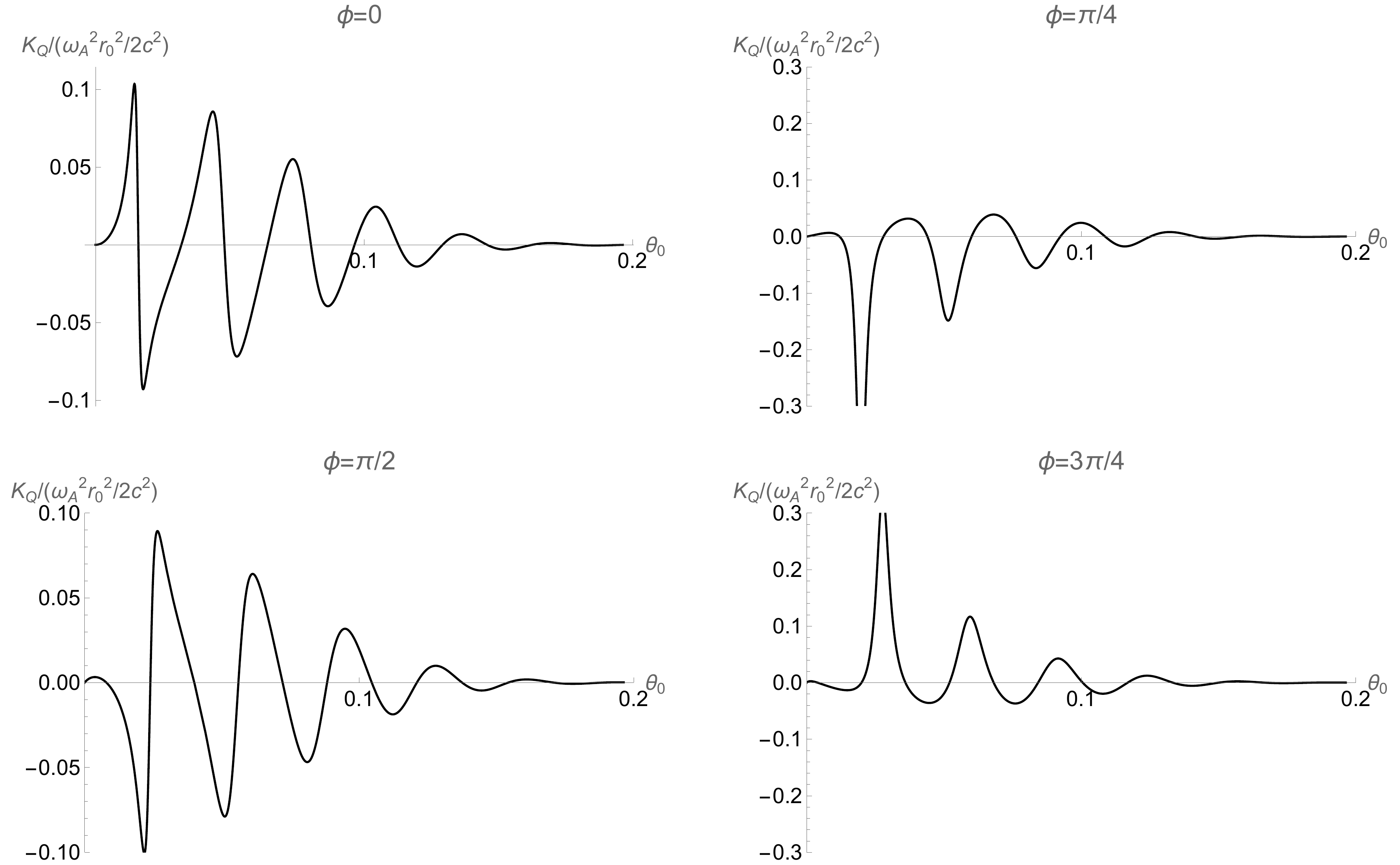}
	\caption{ \label{fig2}
	$K_Q/(\omega_\mathrm{A}^2r_0^2/2c^2)$ as a function of $\theta_0$ 
	for several $\phi$.   
	 } 
\end{figure}

To make the effect of quantum time dilation manifest, 
as in \cite{Chiba:2022fzo} we split the time 
dilation formula (\ref{quantum:dilation}) into $K_C$ and $K_Q$ as  
$\la T_\mathrm{A}\ra=\tau_\mathrm{B}(1-K_{C}-K_Q)$.  $K_C$ is given by 
 the contribution of a statistical mixture of the coherent states of 
 clock~$\mathrm{A}$ and clock~$\mathrm{B}$, and $K_Q$ is the term due to the 
 interference effect 
 \beqa
 K_C&=&\frac{\omega_\mathrm{A}^2r_0^2}{2c^2}+
\frac{\hbar\omega_\mathrm{A}}{2m_\mathrm{A}c^2}-\frac{\hbar^2}{2m_\mathrm{B}^2c^2\sigma_\mathrm{B}^2},\\
K_Q&=&2\sin\theta_0
\frac{\omega_\mathrm{A}^2r_0^2}{Nc^2}{\rm Re}\left(ie^{i(\theta_0-\phi)}e^{\alpha^2-|\alpha|^2}\right)\,.
 \eeqa
Positive $K_Q$ implies the enhanced time dilation. 
In Fig.~\ref{fig2}, $K_Q$ normalized by the classical time dilation 
factor $\omega_\mathrm{A}^2r_0^2/2c^2$ is shown. In this example, 
 the charged clock particle is supposed to be $~^{40}{\rm Ar}^{13+}$ as in 
\cite{King:2022gtf}, and we assumed  
$q_\mathrm{A}=13 e$, $m_\mathrm{A}=6.6 \times 10^{-26}{\rm kg}$, $B= 1.0{\rm T}$ and $r_0=1.0\times 10^{-7}{\rm m}$, so that the classical time dilation factor becomes
 $\omega_\mathrm{A}^2r_0^2/2c^2=5.5 \times 10^{-17}$. 
The quantum effect can either  enhance or reduce the time dilation and can be as large as $10\%$ of the classical time dilation.  The coherence time  
of several seconds for maintaining the superposition may be required 
to observe a quantum time dilation effect, which is an experimental challenge 
but is well within the measurement capability of state-of-the-art clocks 
\cite{2015Natur.528..530K}.

\section{Summary}

As an extension of the proper time observable proposed in \cite{Smith:2019imm} 
and applied to a weak gravitational field \cite{Chiba:2022fzo},  
we studied charged 
quantum clocks interacting with the external  electromagnetic fields. 
We derived a formula of the average proper time read by one 
clock conditioned on another clock reading a different proper 
time, Eq. (\ref{timedelay:final}), which has the same form as that in classical relativity consisting of 
kinetic part (velocity squared term) and gravitational 
part (gravitational redshift term).  
We found that the time dilation is given by difference of velocity 
and distance between trajectories of each clock, regardless 
of whether the clock is in inertial motion or non-inertial motion.

When applied to a charged quantum clock in a uniform magnetic field,  
we considered the case in which the state of one clock is in a 
superposition. 
We found that the effect arising from quantum interference appears in the time dilation which can be as large as $10\%$ of the classical time dilation.  

According to the proper time observable, the time dilation is given by the expectation value depending on how
one prepared clock particle states as in Eq.  (\ref{timedelay:final}). In this paper, to analytically estimate deviation from the classical
time dilation on the basis of the derived formula, we have considered the simplest clock model and have 
employed the coherent states which follow trajectories of
semi-classical cyclotron motion. However, adopting other states or settings, such as
eigenstates of the Hamiltonian and so on, may make it more advantageous to experimentally implement within reach
of currently established technologies. For example, Bushev et al. \cite{Bushev:2016cha} have proposed an experiment with a single electron
in a Penning trap to probe the time dilation depending on the radial cyclotron state of the electron by using the
electronic spin precession as an internal clock.

Optical clocks based on highly charged ions have been considered as 
 a new class of references for highest-accuracy clocks and 
 precision tests of fundamental physics \cite{Kozlov:2018mbp}. 
 Moreover, such an optical  clock based on a highly charged ion 
 was realized recently \cite{King:2022gtf}. 
 Our study may be relevant in interpreting the measurements 
 of the time dilation of a highly charged optical clock.

\section*{Acknowledgments}
This work is supported by JSPS Grant-in-Aid for Scientific Research Number 
22K03640 (TC), Nos.~16K17704 and 21H05186 (SK), and in part by Nihon University. 


\appendix

\section{Quantum Mechanics of a Charged Particle in a Uniform Magnetic Field}
\label{appendixA}

Here, we summerize the basic results on quantum mechanics of a 
charged particle in a uniform magnetic field \cite{Landau,Schulman}

\subsection{Hamiltonian and Relative Coordinate}

Consider a particle with the mass $m$ and the charge $q$ moving 
in a uniform magnetic field $B$. Take 
the $z$-axis in the direction of the magnetic field and assume that the particle 
moves in the $xy$-plane.  

The Hamiltonian  in the symmetric gauge
\beqa
\A=\frac12 \B\times \x=\frac{B}{2}(-y,x,0)
\label{gauge:symmetric}
\eeqa
is given by
\beqa
H=\frac{({\bf p}-q\A)^2}{2m}=\frac{1}{2m}\left(p_x+\frac{m\omega}{2}y\right)^2+
\frac{1}{2m}\left(p_y-\frac{m\omega}{2}x\right)^2\,,
\eeqa
where we have introduced the cyclotron frequency $\omega=qB/m$. 

{}Since  the time evolution of position operator is given from the Heisenberg equation by 
$\dot x_i=\frac{[x_i,H]}{i\hbar}=\frac{p_i-qA_i}{m}$, 
 considering the classical cyclotron motion, we introduce 
the position operators $X$ and $Y$ corresponding to 
the center of the circle 
\begin{equation}
 X=\frac{p_y+m\omega x/2}{m\omega},\quad Y=-\frac{p_x-m\omega y/2}{m\omega}\,,
\end{equation}
and 
the operators $\xi$ and $\eta$ corresponding to the relative coordinates
\beqa
\xi=x-X=-\frac{p_y-m\omega x/2}{m\omega},~~~~~\eta=y-Y=\frac{p_x+m\omega y/2}{m\omega}\,.
\label{relative}
\eeqa
Note that 
both $X$ and $Y$ commute with the Hamiltonian, $[X,H]=0=[Y,H]$, and hence 
they are conserved, but 
$X$ and  $Y$ do not commute with each other, 
$[X,Y]=-i\hbar /m\omega $. 

\subsection{Creation and Annihilation Operators}

We introduce the following creation and annihilation operators 
\beqa
a&=&\sqrt{\frac{m\omega}{2\hbar }}\left(\xi+i\eta\right)=
\sqrt{\frac{m\omega}{2\hbar }}\left(\left(\frac{x}{2}+i\frac{p_x}{m\omega}\right)
+i\left(\frac{y}{2}+i\frac{p_y}{m\omega}\right)
\right),
\label{annihilation:a}\\
a^\dagger &=&\sqrt{\frac{m\omega}{2\hbar }}\left(\xi-i\eta\right)=
\sqrt{\frac{m\omega}{2\hbar }}\left(\left(\frac{x}{2}-i\frac{p_x}{m\omega}\right)
-i\left(\frac{y}{2}-i\frac{p_y}{m\omega}\right)
\right),\label{creation:a}\\
b&=&\sqrt{\frac{m\omega}{2\hbar }}\left(X-iY\right)=
\sqrt{\frac{m\omega}{2\hbar }}\left(\left(\frac{x}{2}+i\frac{p_x}{m\omega}\right)
-i\left(\frac{y}{2}+i\frac{p_y}{m\omega}\right)
\right),\label{annihilation:b}\\
b^\dagger&=&\sqrt{\frac{m\omega}{2\hbar }}\left(X+iY\right)=
\sqrt{\frac{m\omega}{2\hbar }}\left(\left(\frac{x}{2}-i\frac{p_x}{m\omega}\right)
+i\left(\frac{y}{2}-i\frac{p_y}{m\omega}\right)
\right),\label{creation:b}\, 
\eeqa
where $a$ and $b$ commute with each other and obey the usual commutation relations
\beqa
[a,a^{\dagger}]=1,~~~~~[b,b^{\dagger}]=1\,.
\eeqa 
Then, the Hamiltonian and the $z$ component of the angular 
momentum $L_z$ are written in terms of $a$ and $b$  in simple form as
\beqa
H&=&\hbar \omega\left(a^\dagger a+\frac12\right) ,\\
L_z&=&xp_y-y p_x=\hbar (-a^\dagger a+b^\dagger b)\,.
\eeqa
{}From Eqs.~(\ref{annihilation:a})-(\ref{creation:b}), 
the number operator $a^\dagger a$ corresponds to 
the squared distance  from the center of the circle and $b^\dagger b$ corresponds to the squared distance of the center from the origin of the coordinates. 

We also note that 
the center of the circle and the relative coordinates are written 
in terms of creation and annihilation operators as
\beqa
X&=&\frac12 \sqrt{\frac{2\hbar}{m\omega}}(b+b^\dagger),~~~~~Y=\frac{i}{2}\sqrt{\frac{2\hbar}{m\omega}}(b-b^\dagger),\\
\xi&=&\frac12\sqrt{\frac{2\hbar}{m\omega}}(a+a^\dagger),~~~~~
\eta=\frac{i}{2}\sqrt{\frac{2\hbar}{m\omega}}(-a+a^\dagger).
\label{xieta:ab}
\eeqa

\subsection{Coherent State}

As in the case of one-dimensional harmonic oscillator, we introduce 
the coherent state $|\alpha,\beta\rangle$ such that $a|\alpha,\beta\rangle=\alpha|\alpha,\beta\rangle$ and 
$b|\alpha,\beta\rangle=\beta|\alpha,\beta\rangle$, 
which is constructed by applying the operators $e^{\alpha a^{\dagger}}$ and $e^{\beta b^{\dagger}}$ on 
the ground state $|0\rangle$ as
\beqa
|\alpha,\beta\rangle=e^{-\frac{|\alpha|^2+|\beta|^2}{2}}e^{\alpha a^{\dagger}}
e^{\beta b^{\dagger}}|0\rangle\,.
\label{coherentstate}
\eeqa
Then,  from Eq. (\ref{annihilation:a}) and Eq. (\ref{annihilation:b}),  
the eigenvalues 
$\alpha$ and $\beta$ corresponding to  the relative coordinate 
$(r_0\cos\theta_0,r_0\sin\theta_0)$ and 
the center of the circle $(X_0,Y_0)$ are given by
\beqa
\alpha&=&\sqrt{\frac{m\omega}{2\hbar }}r_0e^{i\theta_0},\\
\beta&=&\sqrt{\frac{m\omega}{2\hbar }}(X_0-iY_0) .
\eeqa
The wave function of the coherent state is given by 
\begin{equation}
 \begin{aligned}
  \la \x|\alpha,\beta\ra= &\sqrt{\frac{m\omega}{2\pi\hbar}}
  \exp\left\{
  - \frac{m\omega}{4\hbar}\left[(x - r_0\cos\theta_0 - X_0)^2 
  + (y - r_0\sin\theta_0 - Y_0)^2 \right]
  \right\} \\
  &\times \exp \left\{
  i\frac{m\omega}{2\hbar}
  \left[(r_0\sin\theta_0 - Y_0)x - (r_0\cos\theta_0 - X_0)y
  - r_0 (X_0\sin\theta_0 - Y_0\cos\theta_0)\right]
  \right\} .
 \end{aligned}
\end{equation}
 $a(t)$ and $b(t)$ evolve according to the Heisenberg equation as
\beqa
i\hbar \dot a(t)&=&[a(t),H] =\hbar \omega a(t),\\
i\hbar \dot b(t)&=&[b(t),H] =0\,.
\eeqa
Hence, we have $a(t)=e^{-i\omega t}a$ and $b(t)=b$.   
Then, from Eq. (\ref{xieta:ab}), the expectation values of $\xi(t)$ and $\eta(t)$ 
in the coherent state are given by
\beqa
\langle \xi(t)\rangle&=&\frac12\sqrt{\frac{2\hbar}{m\omega}}
\langle\alpha,\beta|(a(t)+a^\dagger(t))|\alpha,\beta\rangle=\frac12\sqrt{\frac{2\hbar}{m\omega}}(\alpha e^{-i\omega t}+\alpha^*e^{i\omega t})=r_0\cos(\theta_0-\omega t),\label{mean:xi}\\
\langle \eta(t)\rangle&=&\frac{i}{2}\sqrt{\frac{2\hbar}{m\omega}}
\langle\alpha,\beta|(-a(t)+a^\dagger(t))|\alpha,\beta\rangle=\frac12\sqrt{\frac{2\hbar}{m\omega}}(-\alpha e^{-i\omega t}+\alpha^*e^{i\omega t})=r_0\sin(\theta_0-\omega t)\label{mean:eta}\,.
\eeqa
This corresponds to the position of a charged particle orbiting clockwise 
 about the center with the angular velocity $\omega$.\footnote{For a negatively charged particle  
$\omega=qB/m<0$, the particle orbits counterclockwise.} 
The expectation values of $X(t)$ and $Y(t)$ do not depend on time: 
$\langle X(t)\rangle = X_0$ and $\langle Y(t)\rangle = Y_0$.

The expectation value of the Hamiltonian becomes
\beqa
\la \alpha,\beta|H|\alpha,\beta\ra=\hbar\omega \left(|\alpha|^2+\frac12\right)=
\frac12 m\omega^2r_0^2+\frac12\hbar\omega .
\label{hamiltonian:coherent}
\eeqa

\subsection{Time Dilation in a Rotating Frame}

We show that the time dilation Eq. (\ref{timedelay2}) is invariant even 
if we move to a rotating frame. 

Consider a frame $(x',y')$ which rotates with the angular 
velocity $\Omega$ about the $z$ axis with respect the inertial frame $(x,y)$. 
The two coordinates are related by
\beqa
\begin{pmatrix}
x' \\
y' \\
\end{pmatrix}
=
\begin{pmatrix}
\cos\Omega t & \sin\Omega t \\
-\sin\Omega t & \cos\Omega t \\
\end{pmatrix}
\begin{pmatrix}
x \\
y \\
\end{pmatrix}
.
\label{rotation}
\eeqa
Then,  the shift vector appears in the rotating frame
\beqa
-c^2 dt^2+dx^2+dy^2=-c^2dt^2+(dx'-\Omega y' dt)^2+(dy'+\Omega x' dt)^2 ,
\eeqa
that is, $\beta^{x'}c=-\Omega y'$ and $\beta^{y'}c=\Omega x'$. In the presence of 
the shift vector, the (external) Hamiltonian becomes $H=\frac{({\bf P}-q\A)^2}{2m}-\boldsymbol{\beta}c\cdot({\bf P}-q\A)$, so that the time evolution of the 
position vector is given by
\beqa
\dot\x'=\frac{[\x',H]}{i\hbar}=\frac{{\bf P}-q\A}{m}-\boldsymbol{\beta}c\,.
\eeqa
Moreover, from Eq. (\ref{rotation}) and Eq. (\ref{relative}), we have
\beqa
\dot x'&=&\Omega y'+\omega (\eta\cos\Omega t-\xi\sin\Omega t) ,\\
\dot y'&=&-\Omega x'-\omega(\xi\cos\Omega t+\eta\sin\Omega t)\,.
\eeqa
Hence 
\beqa
\frac{({\bf P}-q\A)^2}{m^2}
&=&(\dot x'+\beta^{x'}c)^2+(\dot y'+\beta^{y'}c)^2\nonumber\\
&=&\omega^2\left(\xi^2+\eta^2\right)=\dot x^2+\dot y^2\,.
\eeqa
Therefore,  the time dilation formula Eq. (\ref{timedelay2}) holds 
in a rotating frame. This implies, in particular, that even if we move to a rotating frame with $\Omega=-\omega$ so that a particle is at rest (classically), 
the time dilation does not change. 

\bibliographystyle{apsrev4-1}
\bibliography{references}

\end{document}